\documentclass[prb,twocolumn,showpacs,superscriptaddress]{revtex4}

\usepackage{graphicx}
	\graphicspath{{./img/}}
\usepackage{bm}
\usepackage{amsmath,amssymb}
 
\begin{document}
\title
{Dissipative Landau-Zener transitions of a qubit:
bath-specific and universal behavior}


\author{Keiji Saito}
\affiliation{Department of Physics, Graduate School of Science,
University of Tokyo, Tokyo 113-0033, Japan}
\affiliation{Department of Physics, Graduate School of Science,
2 CREST, JST, 4-1-8 Honcho Kawaguchi, Saitama, 332-0012, Japan}
\author{Martijn Wubs}
\affiliation{Institut f\"{u}r Physik, Universit\"{a}t Augsburg,
Universit\"{a}tsstra{\ss}e 1, D-86135 Augsburg, Germany}

\author{Sigmund Kohler}
\affiliation{Institut f\"{u}r Physik, Universit\"{a}t Augsburg,
Universit\"{a}tsstra{\ss}e 1, D-86135 Augsburg, Germany}

\author{Yosuke Kayanuma}
\affiliation{Department of Mathematical Science, Graduate School of
Engineering, Osaka Prefecture University, Sakai 599-8531, Japan}

\author{Peter H\"{a}nggi}
\affiliation{Institut f\"{u}r Physik, Universit\"{a}t Augsburg,
Universit\"{a}tsstra{\ss}e 1, D-86135 Augsburg, Germany}

\date{\today}

\begin{abstract}
We study Landau-Zener transitions in a qubit coupled to a
bath at zero temperature. 
A general formula is derived that is
applicable to models with a non-degenerate ground state.
We calculate exact transition probabilities 
for a qubit coupled to either
a bosonic or a spin bath. The nature of the baths and the
qubit-bath coupling is reflected in the transition probabilities.
For diagonal coupling, when the bath causes energy
fluctuations of the diabatic qubit states but no transitions between
them, the transition probability coincides with the standard LZ
probability of an isolated qubit.  This result is universal as it
does not depend on the specific type of bath.
For pure off-diagonal coupling, by contrast, the tunneling probability
is sensitive to the coupling strength.
We discuss the relevance of our results for experiments on molecular
nanomagnets, in circuit QED, and for the fast-pulse readout of
superconducting phase qubits.  

\end{abstract}

\maketitle
\section{Introduction}

Nonadiabatic transitions at avoided level crossings play an essential
role in numerous dynamical phenomena in physics and chemistry. They
have been studied both theoretically and experimentally in various
contexts like spin-flip processes in nano-scale magnets,
\cite{Thomas1996a, DeRaedt1997a, Wernsdorfer1999a,Wernsdorfer2006a} 
molecular collisions,
\cite{Child1974a}
optical systems,
\cite{Spreeuw1990a, Bouwmeester1995a}
quantum-dot arrays,
\cite{Saito2004a}
Bose-Einstein condensates,
\cite{Witthaut2006a}
the control of chemical reactions,
\cite{Zhu1994a}
and recently in particular in quantum information processing.
\cite{Ankerhold2003a, Shytov2003a, Izmalkov2004a, Ithier2005a,
Wubs2005a, Shevchenko2005a, Oliver2005a, Saito2006a, Hicke2006a,
Sillanpaa2006b}

The ``standard'' Landau-Zener (LZ) problem describes the ideal situation
in which the dynamics is restricted to
two levels that are coupled by a constant tunnel matrix element and
cross at a constant velocity.   The quantity of primary interest is the
probability that finally the system ends up in the one or the other of
the two states.  This classic problem was solved
independently by several authors in 1932. \cite{Landau1932a, Zener1932a,
Stueckelberg1932a, Mayorana1932a}
In quantum devices, not only the transition probability but
also the nonadiabatic relative phase (Stokes phase) between the two
states is important.\cite{Shytov2003a, Saito2004a, Wubs2005a}
This phase leads to observable interference effects, for
example in superconducting qubits.\cite{Oliver2005a,Sillanpaa2006b}

In an experiment, the two-level system will be influenced by its environment,
which may affect the quantum phase of the superposition, 
alter the effective interaction between the levels, 
or may cause spontaneous decay.
For qubits in a solid-state environment,
\cite{Tian2002a, vanderWal2003a,Simmonds2004a,Cooper2004a,Martinis2005a} 
all these processes may occur simultaneously and hinder qubit manipulation.
Thus in the context of solid-state quantum information processing, a realistic study of 
qubit manipulation via Landau-Zener transitions should include the
influence of environmental degrees of freedom.

The environment of a quantum system can often be described as a bath
of harmonic oscillators.  \cite{Magalinskii1959a, Leggett1987a,
Hanggi1990a, Grifoni1998a, Weiss1999a}  In some situations, it is
known that the dominant environmental effects can best be modelled as
a spin bath instead,\cite{Shao1998a,Prokofev2000a, Hutton2004a} for
example  for molecular magnets\cite{Wernsdorfer1999a,
Wernsdorfer2006a} and for Josephson phase qubits
\cite{Cooper2004a,Simmonds2004a,Martinis2005a} at very low
temperatures. 

In the presence of a heat bath, the Landau-Zener dynamics will
sensitively depend on the qubit operator to which the bath
couples.\cite{Wubs2006a} Ao and Rammer\cite{Ao1989a, Ao1991a} studied
the LZ problem for the special case in which an ohmic heat bath
couples to the same operator as the driving and derived the transition
probabilities in the limit of high and of low temperatures. In the
limits of very fast and very slow sweeps at zero temperature, they
found that the transition probability is the same as in the absence of
the heat bath, as was confirmed by numerical
studies.\cite{Kayanuma1998a, Kobayashi1999a}

This zero-temperature result was recently proven to hold exactly for
{\em arbitrary} Landau-Zener sweep speeds, as a special case of an
exact expression for arbitrary qubit-bath couplings and spectral
densities.\cite{Wubs2006a} Very recently, Pokrovsky and
Sun\cite{Pokrovsky2007a} developed an interesting finite-temperature
formalism valid for baths with short correlation times, in which the
exact transition probabilities of Ref.~\onlinecite{Wubs2006a} indeed
show up in the zero-temperature limit.
At sufficiently high temperatures, the qubit experiences essentially
classical Gaussian white noise and the Landau-Zener problem can be
solved exactly.\cite{Kayanuma1984a,Saito2002a}

A qubit that undergoes a Landau-Zener sweep while coupled to another
quantum system, is equivalent to a multi-level Landau-Zener problem in
which two groups of levels cross.  If all avoided crossings are
sufficiently well separated, the dynamics consists of effectively
independent transitions and one can compute the transition
probabilities by transfer-matrix techniques.  \cite{Kayanuma1993a,
Saito2004a, Kayanuma2000a, Wubs2005a}
Independent level crossings do not occur for the dissipative
Landau-Zener problem, for which the adiabatic energy spectrum is
continuous. Other methods are therefore required, for example the
exact summation of a perturbation series.\cite{Kayanuma1984a}

Brundobler and Elser\cite{Brundobler1993a} considered a special
multi-level Landau-Zener problem in which the system starts out in the
diabatic state whose energy changes faster than that of all other
diabatic levels.  They conjectured that the transition
probabilities are then given by an expression that only contains the
velocities of the diabatic levels and off-diagonal matrix elements of
the Hamiltonian. The conjecture has long known to be true for the
exactly solvable Demkov-Osherov model, where {\em one} level crosses
several parallel levels.\cite{Demkov1968a, Brundobler1993a} 
Analytical proofs of the Brundobler-Elser conjecture have
been given recently for several multi-level crossing situations.
\cite{Shytov2004a,Volkov2004a,Dobrescu2006a,Volkov2006a,Ostrovsky2006a}

The physically important situation of a two-level system coupled to a
heat bath corresponds to two continuous bands of diabatic levels, with
the same energy-level velocities within each band. A generalization
of the theorem by Brundobler and Elser to this case will be considered. 
This general formula applies to dissipative Landau-Zener problems
for all those kinds of baths at zero temperature for which the initial
qubit-plus-bath ground state is unique. We discuss the implications
of our results for a wide range of experiments for which these
conditions are met.  As one of our main results, we present a
universal aspect of dissipative Landau-Zener transitions, independent
of the precise nature of the bath. 

The paper is organized as follows:
In Sec.~\ref{sec:model} we review the standard LZ formula for a
two-level system and generalize it to the multi-level case;
the explicit calculations have been deferred to the Appendix.
Section \ref{sec:bath} is devoted to the derivation of the LZ
transition probability for both a harmonic oscillator bath and a spin
bath. In Sec.~\ref{sec:nonlinear}, we identify a universality 
in Landau-Zener tunneling that holds true even for baths
of nonlinear oscillators. Finally, we discuss several promising
applications and experiments in Sec.~\ref{sec:experiments}.

\section{Landau-Zener tunneling probabilities}
\label{sec:model}
 
\subsection{Landau-Zener transitions in an isolated qubit}

To set the stage and to introduce our notation, we first review
the standard Landau-Zener problem for an isolated qubit.  By 
``isolated'' we mean: not coupled to an
environment but nevertheless driven by a deterministic classical field, 
whose physical origin will be specified and discussed in Sec.~\ref{sec:experiments}. The two-level Hamiltonian reads
\begin{equation} \label{LZ_isolated} 
\mathcal{H}_{\rm LZ}(t)
= \frac{vt}{2}{\bm \sigma}_{z} + \frac{\Delta}{2}{\bm \sigma}_{x}, 
\end{equation}
where ${\bm {\bm \sigma}}_{z} = |{\uparrow}\rangle\langle{\uparrow}|
-|{\downarrow}\rangle\langle{\downarrow}|$ and 
${\bm {\bm \sigma}}_{x} = |{\uparrow}\rangle\langle{\downarrow}|
+|{\downarrow}\rangle\langle{\uparrow}|$ are Pauli matrices while
$|{\uparrow}\rangle$ and $|{\downarrow}\rangle$ denote the so-called
diabatic states with the energies $\pm \frac{1}{2}vt$ which cross at
$t=0$.
Two parameters determine the dynamics: the constant sweep velocity
$v>0$ by which the energies of the diabatic states cross, and the coupling
matrix element $\Delta$ between these states. Without loss of
generality, we assume $\Delta$ to be real and non-negative.  For
$\Delta\neq 0$, the diabatic states are not eigenstates of the
Hamiltonian~\eqref{LZ_isolated}, so that generally a population
transfer is induced. The Hamiltonian~\eqref{LZ_isolated} is
time-dependent and so are its (adiabatic) eigenstates.  In the limit
$|t|\gg \Delta/v$, the adiabatic eigenstates coincide with the
diabatic states.
\begin{figure}[t]
\includegraphics{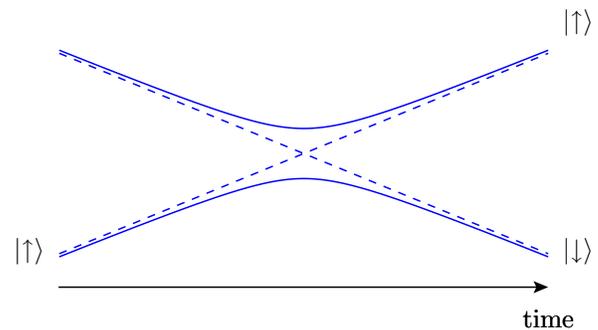}
\caption{(Color online) Adiabatic (solid) and diabatic (dashed) energies for the
standard Landau-Zener problem.}
\label{fig:energies}
\end{figure}

The diabatic energies cross, but the adiabatic energies
$\pm \frac{1}{2}\sqrt{v^{2}t^{2}+\Delta^{2}}$  for
$\Delta\neq 0$ form an avoided crossing, as sketched in
Fig.~\ref{fig:energies}. The adiabatic theorem \cite{Kato1950a} states
that the splitting $\Delta$ prevents  transfer of population between
the adiabatic eigenstates in the adiabatic limit $\hbar v \ll \Delta^{2}$, in
other words if the sweep occurs slowly enough. A qubit prepared
at $t=-\infty$ in the initial ground state $|{\uparrow}\rangle$ will
then end up in the final ground state $|{\downarrow}\rangle$. Beyond the adiabatic regime, 
the dynamics can be rather complex.  Nevertheless, the population of the diabatic states
at $t=\infty$ can be calculated exactly and is determined by the
Landau-Zener transition probability \cite{Landau1932a, Zener1932a,
Stueckelberg1932a}
\begin{equation}\label{PLZ_isolated}
P_{{\uparrow \rightarrow \downarrow}}
\equiv \big|\langle\psi(\infty)|{\downarrow}\rangle \big|^{2}
=      1 - \exp\Big(-\frac{\pi \Delta^{2}}{2\hbar v}\Big) ,
\end{equation}
which denotes the probability for a \textit{transition} to the
opposite diabatic state, i.e.\ the probability to \textit{stay} in the
adiabatic eigenstate. Accordingly, $P_{{\uparrow \rightarrow \uparrow}}
= 1- P_{{\uparrow}\to{\downarrow}}$ denotes the probability for a
non-adiabatic transition, i.e.\ a jump across the avoided crossing.

\subsection{Landau-Zener transitions in non-isolated qubits}
\label{Secnonisolated}

We now turn to situations where the qubit
is no longer isolated.
The Hamiltonian
of the driven qubit plus its environment has the general form
\begin{eqnarray} \label{LZbath}
\mathcal{H}(t) &=& \mathcal{H}_{\rm LZ}(t) + \mathcal{H}_\text{q-env} +
\mathcal{H}_\text{env},
\end{eqnarray} 
where $\mathcal{H}_{\rm env}$ describes the environment Hamiltonian with
Hilbert space of dimension $M \le \infty$.  We assume the most general
linear coupling between the qubit operators ${\bm {\bm \sigma}}_{x}$,
${\bm {\bm \sigma}}_{y} =-i(|{\uparrow}\rangle\langle{\downarrow}|
-|{\downarrow}\rangle\langle{\uparrow}|)$, ${\bm {\bm \sigma}}_{z}$
and environment operators $\mathcal{X}^{x,y,z}$, 
in other words we take the qubit-environment coupling 
\begin{equation}\label{qbath}
\mathcal{H}_\text{q-env}
= \sum_{\nu=x,y,z}{\bm \sigma}_{\nu} \mathcal{X}^{\nu} .
\end{equation}
We denote by $|k\rangle$ the eigenstates of the environment
Hamiltonian $\mathcal{H}_\text{env}$. 

An important assumption underlying our model~\eqref{LZbath} is that the
qubit-bath coupling~\eqref{qbath} and the bath itself are not affected
by the driving.
Then at very large times $t\to\pm\infty$, the qubit Hamiltonian is
dominated by the time-dependent part, so that all states of the system-plus-environment
belong to one of two bands: an ``up-cluster''
$|{\uparrow}\rangle|k\rangle$ and a ``down-cluster''
$|{\downarrow}\rangle|k\rangle$, with energies moving upwards and
downwards, respectively.  

\subsubsection{Diabatic basis}
The dissipative Landau-Zener problem is a scattering problem
in the restricted sense that changes in the qubit's state will occur only 
during a finite time interval around $t=0$.
In order to exploit this fact that the qubit will not flip for
sufficiently large $|t|$, we decompose $\mathcal{H}(t)$
into its diabatic states. These are the eigenstates of the total
Hamiltonian \eqref{LZbath} in the the limits
$t\to\pm\infty$.  Initially and finally, the Hamiltonian is dominated by the
term proportional to $\bm{\sigma}_z$, so that the diabatic basis 
for the qubit is simply given by the states $|{\uparrow}\rangle$ and
$|{\downarrow}\rangle$.  

\begin{figure}[t]
\includegraphics{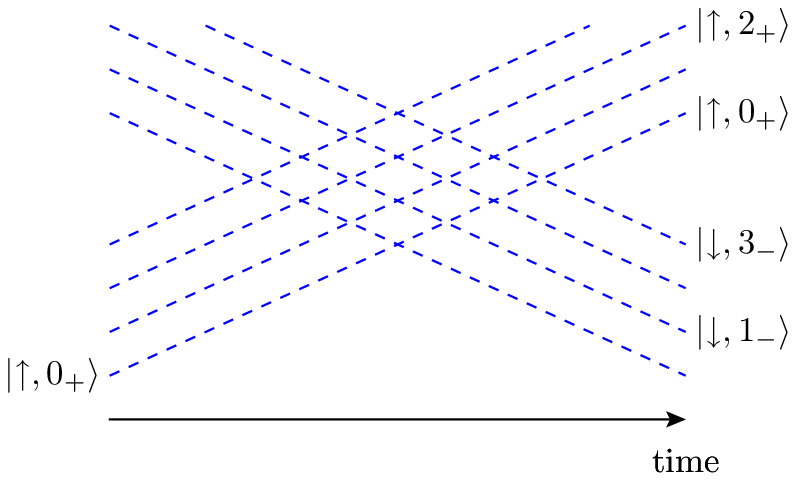}
\caption{(Color online) Sketch of the diabatic energy levels as a function of time for 
a qubit coupled to a {\em single} harmonic oscillator.  Energies of
the states in the ``up cluster'' increase. These states correspond to
the qubit state $|{\uparrow}\rangle$. Energies decrease in the ``down
cluster'', where the qubit state is $|{\downarrow}\rangle$.  According
to the ``no-go-up theorem'' \eqref{app:nogoup}, the initial state
$|{\uparrow},0_+\rangle$ evolves to a superposition in which
$|{\uparrow},0_+\rangle$ is the only ``up'' state. Energies within a
band are separated by the oscillator energy $\hbar\Omega$. For a qubit
coupled to an oscillator bath, the corresponding crossing clusters
would be continuous bands of states.
}
\label{fig:multilevel}
\end{figure}

For the environment, by contrast, there is no corresponding growing
energy scale for large $|t|$. Its diabatic states are influenced by
the coupling to the qubit and depend on the qubit's state. 
For the ``up-cluster'', these diabatic
eigenstates of the environment are those which diagonalize the
Hamiltonian projected to the subspace $|{\uparrow}\rangle$, i.e.\
$\langle{\uparrow}|\mathcal{H}|{\uparrow}\rangle$.  They are
eigenstates of $\mathcal{H}_\mathrm{env}+\mathcal{X}^z$, and we denote
them by $|{k_{+}}\rangle$ and their energies by $\varepsilon_{k_{+}}$.
The diabatic bath states for the ``down-cluster'', $|{k_{-}}\rangle$,
are defined likewise so that
\begin{equation}
(\mathcal{H}_\mathrm{env} \pm \mathcal{X}^z) |{k_{\pm}}\rangle
= \varepsilon_{k\pm} |{k_{\pm}}\rangle .
\end{equation}
The diabatic states of the qubit plus the bath read
\begin{subequations}\label{diagstates}
\begin{align}
|{\uparrow}\,{k_{+}}\rangle \equiv |{\uparrow}\rangle|{k_{+}}\rangle,
\label{upclusterstates}
\\ 
|{\downarrow}\,{k_{-}}\rangle\equiv|{\downarrow}\rangle|{k_{-}}\rangle ,
\label{downclusterstates}
\end{align}
\end{subequations}
where the labels $k_\pm = 0,1,2,\ldots$ are assigned such that the
energies $\varepsilon_{k_\pm}$ are in increasing order, see the sketch
in Figure~\ref{fig:multilevel}.  At asymptotically large times,
$t\to\pm\infty$, the diabatic states diagonalize the total Hamiltonian
\eqref{LZbath} and hence coincide with adiabatic eigenstates which
diagonalize $\mathcal{H}(t)$ at a given time $t$.  Note that $\langle
{\downarrow} k_{-}|{\uparrow}{k_{+}}\rangle = 0$, although in general
$\langle{k_{+}}|{{k'}_{-}}\rangle \ne \delta_{k k'}$.
A state of particular interest is the adiabatic ground state
$|{\uparrow}\,{0_{+}}\rangle$ which has energy $(v t /2 +
\varepsilon_{0_{+}})$. At zero temperature, it is the natural initial
state for the Landau-Zener dynamics.

We now split the Hamiltonian \eqref{LZbath} into two parts, one that
is diagonal in the spin index, while the other is off-diagonal. The former part
consists of all terms proportional to $\bm\sigma_z$ and is diagonal
 in the diabatic basis \eqref{diagstates}.
The latter part  reads
\begin{equation}\label{BFinteraction}
\mathcal{V} =  \frac{\Delta}{2} {\bm \sigma}_{x}
+ \bm\sigma_x \mathcal{X}^x + \bm\sigma_y \mathcal{X}^y
\end{equation}
and will be called  the {\em bit-flip interaction}, since it contains
all interaction terms of the Hamiltonian \eqref{LZbath} that induce a
change in the state of the qubit.

An important feature of the diabatic basis \eqref{diagstates} is that
all matrix elements of $\mathcal{V}$ vanish within each cluster, i.e.
\begin{equation}
\langle{\uparrow}k_+|\mathcal{V}|{\uparrow}k_+'\rangle =
\langle{\downarrow}k_-|\mathcal{V}|{\downarrow}k_-'\rangle = 0 .
\end{equation}
This relation will be essential for the application of the general formula for
the nonadiabatic transition probabilities as derived in Appendix
\ref{app:BEgen}.

\subsubsection{Nonadiabatic transition probabilities}
\label{LZbathprob}

We have now achieved a useful formulation of the dissipative
Landau-Zener problem in terms of two groups of diabatic states.  If
the group of upward moving parallel levels would consist of merely one
state, then transition probabilities could be computed with the simple
independent-crossing formula, for which Brundobler and
Elser\cite{Brundobler1993a} conjectured that it holds even when
successive level crossings are not independent. Recent proofs show
that the independent-crossing formula indeed holds exactly, even in
more general situations.\cite{Shytov2004a, Volkov2004a,
Dobrescu2006a,Volkov2006a, Ostrovsky2006a} As stated above, for
dissipative Landau-Zener transitions there are two continua of states
that cross with constant velocity.  This physically important
situation is addressed in the Appendix~\ref{app:BEgen}, where exact
nonadiabatic transition probabilities are derived in a fairly general
setting, with the crossing of two continua of parallel states as a
special case. This setting is a generalization of our recent studies
in Refs.~\onlinecite{Saito2006a} and \onlinecite{Wubs2006a}.

For the dissipative Landau-Zener problem, we can deduce the following
from Eqs.~\eqref{app:nogoup} and \eqref{app:Paa.v}: If at $t=-\infty$
the system starts  in a state $|{\uparrow}k_+\rangle$ whose diabatic
energy is non-degenerate, then the following transition probabilities
at $t=\infty$ are exact:
\begin{equation}
P_{{\uparrow k_{+}} \to {\uparrow k'_{+}}} =
\begin{cases}
\exp\left(-\frac{2\pi \langle {\uparrow k_{+}}|\,\mathcal{V}^{2}
           |{\uparrow k_{+}}\rangle}{ \hbar v}\right)
  & \text{for $k_+'=k_+$} ,
\\
0 & \text{for $k_+'> k_+$} .
\end{cases}
\label{formula_2}
\end{equation}
For the transition to lower states within the initial band of states
($k_+'< k_+$), we cannot make any statement.  The second line of
Eq.~\eqref{formula_2} asserts that states of the ``up-cluster'' that
lie above the initial state will finally be unpopulated. This no-go
theorem was formulated in Refs.~\onlinecite{Sinitsyn2004a} and
\onlinecite{Volkov2005a}, and we think that it is more aptly described by
the name ``no-go-up theorem''.

A case of particular interest is that of the initial state
$|{\uparrow}0_+\rangle$, which is the ground state of the entire system.
For all bath models employed below, the ground state is unique, so that
relation \eqref{formula_2} applies.
Then final states with $k_+'< k_+$ do not exist, while the occupation
of final states with $k_+'>k_+$ is forbidden by the no-go-up theorem.
Thus, provided that the final qubit state is $|{\uparrow}\rangle$, the
environment will end up in its ground state.

It is the final transition probabilities $P_{\uparrow\rightarrow\uparrow}$ 
and $P_{\uparrow\rightarrow\downarrow}$ for the qubit that interests
us most, irrespective of the final state of the environment. By
tracing out the environment, i.e.\ by performing the sum over $k_+'$,
we find
\begin{equation}
\label{PLZopen}
P_{{\uparrow} \rightarrow {\uparrow}}
= \exp\Big(- \frac{\pi W^{2}}{2\hbar v} \Big)
=
1- P_{{\uparrow} \rightarrow {\downarrow}}
\end{equation}
with the ground-state expectation value
\begin{equation}
\label{PLZopen_W}
W^{2} = 4 \langle {\uparrow} 0_+|\,\mathcal{V}^{2}\,|{\uparrow}
0_{+}\rangle
\end{equation}
These are the two central equations for dissipative Landau-Zener
transitions at zero temperature.  The ground-state expectation value
$W^{2}$ formally replaces squared the tunnel matrix element $\Delta^{2}$ in
the original Landau-Zener expression \eqref{PLZ_isolated}.

\section{Dissipative Landau-Zener transitions in various environments}
\label{sec:bath}

We are now in the position to study Landau-Zener transitions for
specific baths with  linear couplings to the qubit.  We will focus on
the two most important model baths in quantum dissipation, namely a
bath of harmonic oscillators and a spin bath.  
In both cases, we will restrict ourselves to zero temperature, so that
the natural initial state is the diabatic ground state $|{\uparrow}
0_+\rangle$ of the system plus the environment.  Both for the
harmonic-oscillator bath and for the spin bath, the ground state is
non-degenerate, so that formula~\eqref{PLZopen} can be applied.  The
essential steps that remain are first to identify and characterize the
diabatic ground state $|{\uparrow} 0_+\rangle$, and then to compute
the expectation value \eqref{PLZopen_W}.
Applications to specific experiments will be discussed in
Sec.~\ref{sec:experiments}.

\subsection{Harmonic-oscillator bath}\label{sectionbosonicbath}

We first consider the case in which a qubit interacts with a standard
bosonic bath consisting of harmonic oscillators.   The Hamiltonian is
as in Eq.~\eqref{LZbath}, with the environment Hamiltonian
\begin{equation}\label{boson_bath} 
\mathcal{H}_{\rm env}
= \sum_{j = 1}^{N} \hbar \Omega_{j} b_{j}^{\dag}b_{j}
\end{equation}
consisting of $N$ harmonic oscillators with frequencies $\Omega_{j}$.
Zero-point energies do not play a role here and are ignored.
The $b_{j}^{\dagger}$ and $b_{j}$ denote the usual creation and
annihilation operators of the oscillator $j$.  We leave the number of
oscillators $N$ finite at first but eventually take it to infinity in
a continuum limit.  Furthermore, we assume the most general linear
qubit-oscillator coupling
\begin{equation}\label{Hspinboson}
\mathcal{H}_\text{q-env} =  
 \sum_{\nu=x,y,z}{\bm \sigma}_{\nu}
 \sum_{j=1}^{N} \frac{\gamma_{j}}{2} \lambda_{j}^{\nu}( b_{j} + b_{j}^{\dagger}),
\end{equation}
where the second sum specifies the
environment operators $\mathcal{X}^{\nu}$ defined in Eq.~\eqref{qbath}.
Since the coupling~\eqref{Hspinboson} also includes the ${\bm \sigma}_{y}$
interaction, this constitutes a  generalization of the spin-boson
model that we considered in Ref.~\onlinecite{Wubs2006a}.
The parameters $\gamma_j$ determine the coupling strengths, while the
parameters $\lambda_j^\nu$ define the ``coupling directions'' and are
conveniently expressed by the spherical coordinates $\vartheta_j$ and
$\varphi_j$ as
\begin{subequations}
\begin{align}
\lambda_{j}^{x} ={}& \sin\vartheta_{j}\cos \varphi_{j} ,\\
\lambda_{j}^{y} ={}& \sin\vartheta_{j}\sin \varphi_{j} , \\
\lambda_{j}^{z} ={}& \cos\vartheta_{j} .
\end{align}
\end{subequations}
The bit-flip interaction \eqref{BFinteraction} then  becomes
\begin{equation}\label{BFosc}
\mathcal{V}
= \frac{\Delta}{2}{\bm \sigma}_{x}
 + \sum_{j = 1}^{N}\frac{\gamma_{j}\sin \vartheta_{j}}{2}
   \left( \bm\sigma_x \cos\varphi_{j}
         +\bm\sigma_y \sin \varphi_{j}\right) ( b_{j}+ b_{j}^{\dag}).
\end{equation} 

The diabatic states of the environment are determined by diagonalizing
the Hamiltonian
\begin{align}
\mathcal{H}_\text{env}+\mathcal{X}^z
={}& \sum_j \hbar\Omega_j b_j^\dag b_j
     + \sum_j \frac{\gamma_j}{2}\cos\vartheta_j (b_j^\dag + b_j)
\\
={}& \sum_j \hbar\Omega_j b_{j_+}^\dag b_{j_+}
     -\sum_j E_j . \label{Hupup.shifted}
\end{align}
In order to obtain the diagonal form in the last line, we introduced
the shifted annihilation operators
\begin{equation}
b_{j_+} = b_j + \frac{\gamma_j\cos\vartheta_j}{2\hbar\Omega_j}
\end{equation}
and the reorganization energy of the $j$th oscillator, $E_j =
(\gamma^2/4\hbar\Omega_j)\cos^2\vartheta_j$.
The latter quantity denotes the energy shift of the oscillator
ground state owing to the coupling to the qubit.
From Eq.~\eqref{Hupup.shifted}, it becomes immediately clear that the
diabatic ground state has to fulfill the relation
\begin{equation}
\label{bj.shifted}
b_{j_+} |{\uparrow}0_+\rangle = 0 ,
\end{equation}
while the excited diabatic states are created by applying the operators
$b_{j_+}^\dag$ to this diabatic ground state.

We now write the bit-flip interaction $\mathcal{V}$ with the shifted
operators $b_{j_+}$ and employ relation \eqref{bj.shifted} to evaluate
the ground state expectation value \eqref{PLZopen_W}.
After some algebra, we obtain\cite{Wubs2006a}
\begin{equation}
W^{2}
= \bigg| \Delta - \sum_j \frac{\gamma_{j}^2  \sin (2
  \vartheta_{j}) e^{-i\varphi_{j}}}{ 2\hbar\Omega_{j} } \bigg|^2 
 +\sum_j \gamma_j^2 \sin^2 \vartheta_{j} ,
\label{n_expression}
\end{equation}
which allows one to compute the Landau-Zener transition
probabilities~\eqref{PLZopen}.  Note the $\varphi$-dependence,
generalizing our recent work,\cite{Wubs2006a} which
shows that for $\Delta\neq 0$, it makes a difference for Landau-Zener
transition probabilities what types of off-diagonal qubit-oscillator
couplings exist. 

\subsubsection{Identical coupling angles}

For system-bath models, it is frequently assumed that all bath
oscillators couple to the central quantum system via the same
coordinate.  In our model, this corresponds to the case in
which all coupling angles are identical, $\vartheta_j=\vartheta$ and
$\varphi_j=\varphi$.  Then Eq.~\eqref{n_expression} becomes
\begin{equation}\label{Wthetaphi}
W^{2}(\vartheta , \varphi )
= \big| \Delta - \frac{1}{2}E_{0}\sin(2\vartheta)e^{- i\varphi} \big|^{2}
  + S\sin^{2}\vartheta,
\end{equation}
in terms of the integrated spectral density\cite{typos}
\begin{equation}
S = \sum_{j}\gamma_{j}^{2}
  = \frac{\hbar^{2}}{4\pi}\int_{0}^{\infty} d\omega\,J(\omega)
    \label{Sdef}
\end{equation}
and the energy  
\begin{equation}
E_{0} = \sum_{j}\frac{\gamma_{j}^{2}}{\hbar \Omega_{j}}
  = \frac{\hbar}{4\pi}\int_{0}^{\infty} d\omega\,
    \frac{J(\omega)}{\omega} ,
    \label{Edef}
\end{equation}
which equals four times the total reorganization energy $E=\sum_{j}E_{j}$.
These quantities $S$ and $E_{0}$ were presented as frequency integrals over the 
spectral density $J(\omega)$ of the bath, with the latter defined as
\begin{equation}
J (\omega )
= \pi\sum_{j=1}^{N} \Big( \frac{2\gamma_{j}}{\hbar}\Big)^{2}
  \delta (\omega - \Omega_{j }).
\label{definition_of_spectr_b}
\end{equation}
In a continuum limit, the spectral density becomes a smooth function of
frequency.  At low frequencies, one typically observes
 a power-law behavior for $J(\omega)$, whereas the qubit becomes insensitive to  
 high frequencies, as characterized by 
 a cutoff frequency $\omega_\mathrm{c}$. An important class of 
spectral densities is therefore \cite{Leggett1987a,Weiss1999a}
\begin{equation}
\label{Js}
J (\omega )= \alpha \omega^s e^{-\omega /\omega_{\rm c} } .
\end{equation}
For such spectral densities,  one immediately obtains the two 
global quantities whereby the bath influences LZ transition probabilities:
\begin{align} 
S ={}& \frac{\alpha\hbar^2}{4\pi} \omega_\mathrm{c}^{s+1}\Gamma(s+1) ,
\\
E_0 ={}& \frac{\alpha\hbar}{4\pi} \omega_\mathrm{c}^s \Gamma(s),
\end{align}
where $\Gamma(x)$ denotes the Euler Gamma function.

\begin{figure}[t]
\centerline{\includegraphics{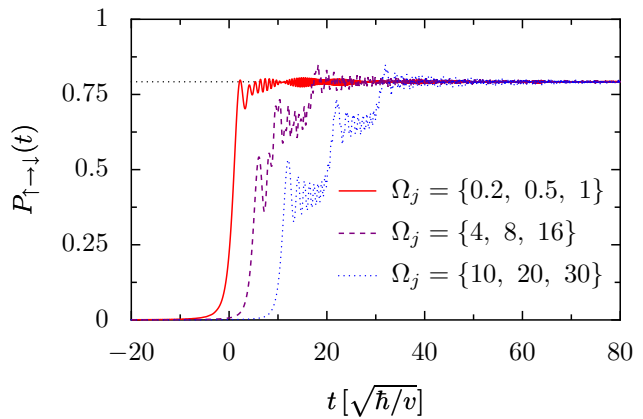}}
\caption{(Color online) Landau-Zener dynamics for a qubit with $\Delta=0$, in all
three cases shown off-diagonally coupled via $\sigma_{x}$ to three
oscillators. The various oscillator  frequencies $\Omega_{j}$ are
given in units of $\sqrt{v/\hbar}$. All coupling strengths have the
same value $\gamma_{j}= \sqrt{\hbar v/3}$. The dotted line marks the
analytical final transition probability corresponding to
Eq.~\eqref{transverse_result}.}
\label{fig:three_osc}
\end{figure}

\paragraph{Off-diagonal coupling.}
For $\vartheta={\pi/2}$, the qubit interacts via its
off-diagonal operators ${\bm\sigma}_{x}$ and $\bm\sigma_y$ with the
environment, whereas the LZ driving affects the qubit only via ${\bm
\sigma}_{z}$.  Equation~\eqref{Wthetaphi} becomes
\begin{eqnarray}
W^{2}(\pi/2,\varphi) = \Delta^2 +S. \label{transverse_result}
\end{eqnarray}
Interestingly enough, the Landau-Zener tunneling probability is then fully
determined by the integrated spectral density $S$.  In particular,
there is no dependence on the oscillator frequencies $\Omega_{j}$.
This result is nicely illustrated in the simple example of
Figure~\ref{fig:three_osc}, showing Landau-Zener dynamics of a qubit
that is coupled to only three oscillators.  The oscillator
frequencies are varied, while the qubit-oscillator couplings are kept
constant.  The dynamics at intermediate times depends on the
oscillator frequencies, but the final transition probability does not.

Note that  $W^{2}(\pi/2,\varphi)$ in Eq.~\eqref{transverse_result} is
independent of $\varphi$, so that in case of {\em only off-diagonal
coupling}   the relative weight of the interactions via $\bm \sigma_x$
and $\bm\sigma_y$ drops out of the final occupation probability.
Since $S>0$, it is clear from Eqs.~\eqref{PLZopen_W} and
\eqref{transverse_result} that a off-diagonal coupling always enhances
the occupation of the final ground state $|{\downarrow}\rangle$ as
compared to the case without dissipation.  This has an intuitive
explanation: the zero-temperature oscillator bath partially succeeds
in  cooling the qubit down to its instantaneous ground state at any
time during the level crossing.

\paragraph{Diagonal coupling.}
For $\vartheta=0$, the qubit interacts with the environment through the
diagonal Pauli matrix ${\bm \sigma}_z$.
This interaction induces pure dephasing
between the states $|{\uparrow}\rangle$ and $|{\downarrow}\rangle$.
This driven spin-boson model with diagonal interaction has so far been
the standard model for discussing Landau-Zener transitions in
dissipative environments.\cite{Kayanuma1984a, Ao1989a, Shimshoni1991a,
Saito2002a, Kayanuma1998a}
For diagonal coupling, the bit-flip interaction \eqref{BFosc} simply
becomes $\mathcal{V}=(\Delta/2)\bm\sigma_x$, so that
\begin{equation}
W(0,\varphi)^2 = \Delta^2 ,
\end{equation}
and the Landau-Zener transition probability~\eqref{PLZopen} coincides
with the standard expression Eq.~\eqref{PLZ_isolated} for an {\em
isolated} qubit.  This bath independence of the transition
probability for diagonal coupling was predicted by Ao and Rammer
in the limits of fast and slow Landau-Zener sweeps.\cite{Ao1989a} Here
we find that at zero temperature, it holds exactly for all
diagonally coupled harmonic-oscillator baths, and for all coupling
strengths $\Delta$ and sweep velocities $v$.\cite{Wubs2006a}

\begin{figure}[t]
\centerline{\includegraphics{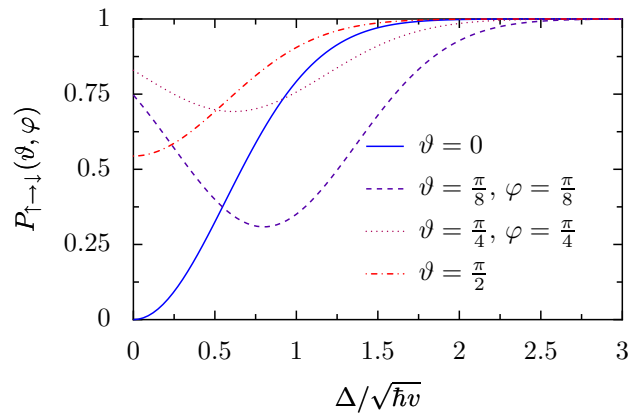}}
\caption{(Color online) Transition probability $P_{\uparrow\to\downarrow}$ as a
function of the internal coupling $\Delta$ for $E_{0}=2\sqrt{\hbar v}$
and $S=0.5\hbar v$.  The angles $\vartheta=0$ and $\vartheta= \pi/2$
correspond to purely diagonal and off-diagonal coupling,
respectively, for which the probability is independent of $\varphi$.
}
\label{fig2}
\end{figure}

\paragraph{General coupling.}
When the oscillators neither couple purely off-diagonally
($\vartheta=\pi/2$) nor purely diagonally ($\vartheta=0$),  the
Landau-Zener probability generally exhibits a non-monotonic dependence
on the tunnel coupling $\Delta$. This is shown in Figure~\ref{fig2}
for various angles $\varphi$ and $\vartheta$.  Most interesting is the
comparison to the non-dissipative case,  which as we saw coincides
with the result for diagonal coupling ($\vartheta=0$): Any dissipative
Landau-Zener probability lower than the curve for $\vartheta=0$ is
counterintuitive.  Such situations occur, however: for several values
of $\vartheta$ and for a sufficiently large tunnel splitting $\Delta$,
the bath coupling reduces $P_{\uparrow\rightarrow\downarrow}(\infty)$,
i.e.\ dissipation enhances the population of the final excited qubit
state.\cite{Wubs2006a}

This counterintuitive behavior is most significant for the angles
$\varphi=0$, $\vartheta = \pi/4$ and when the squared reorganization
energy is large compared to the integrated spectral density $E_0^2\gg
S$. This means that the counterintuitive behavior
would be observable only in rather exceptional situations, which did
not shape our intuition.  For example, for the spectral density
\eqref{Js}, a significant reduction of the ground state population by
increase of $\Delta$ requires a very strong qubit-bath coupling
$\alpha \gtrsim 1$.  In the opposite limit $\alpha \ll 1$, which is
relevant for quantum information processing, the bath for all
practical purposes induces the more intuitive tendency towards the
ground state as $\Delta$ is increased, as we found above for purely
off-diagonal or purely diagonal couplings. 

\subsection{Spin bath}
\label{sec:spinbath}
\setcounter{paragraph}{0}

Let us now turn to the case in which a qubit interacts with an
ensemble of otherwise non-interacting two-level systems forming a spin
bath.\cite{Prokofev2000a, Shao1998a, Hutton2004a}
The total Hamiltonian is again assumed to be of the general
form~\eqref{LZbath}, with the standard Landau-Zener
Hamiltonian~\eqref{LZ_isolated} for the qubit, but now with the bath
Hamiltonian
\begin{equation}
\mathcal{H}_{\rm env} = \sum_{j=1}^{N} \sum_{\nu =x,y,z}
B_{j}^{\nu} {\bm \tau}^{j}_{\nu}, \label{spin_bath}
\end{equation}
where ${\bm \tau}_{\nu}^{j}$ are the Pauli matrices for the $j$th bath
spin.  The most general linear qubit-bath coupling reads
\begin{equation}
\label{intspinbath}
\mathcal{H}_\text{q-env}
= \sum_{\nu =x,y,z} {\bm \sigma}_{\nu}
  \sum_{j=1}^{N} \gamma_{j}^{\nu} {\bm \tau}_{\nu}^{j} ,
\end{equation}
where the second sum defines the operators $\mathcal{X}^\nu$ as a
linear combination of the spin operators ${\bm \tau}^{j}_{\nu}$, with
coupling constants $\gamma_{j}^{\nu}$. The bit-flip interaction
$\mathcal{V}$ then becomes
\begin{equation}\label{BFspinbath}
\mathcal{V}
= \frac{\Delta}{2}{\bm \sigma}_{x} + \sum_{j = 1}^{N}
  \left( \gamma_{j}^{x} {\bm \sigma}_{x} {\bm \tau}^{j}_{x}
 +\gamma_{j}^{y}{\bm \sigma}_{y}  {\bm \tau}^{j}_{y} \right).
\end{equation} 

As for the bosonic bath in Sec.~\ref{sectionbosonicbath}, we wish to
apply the general formula~\eqref{PLZopen} for the transition
probability. For that, we need to determine the diabatic eigenstates of the
qubit plus the spin bath.  For large $|t|$, the time-dependent term
$vt\bm\sigma_z/2$ dominates and, therefore, provides the diabatic
qubit states $|{\uparrow}\rangle$ and $|{\downarrow}\rangle$.
Consequently, the diabatic spin-bath states are determined by the
operator
\begin{equation}
\langle s|\mathcal{H}(t)|s\rangle
= \sum_j \mathcal{H}_{\text{spin},j}^\pm
,\quad s={\uparrow},{\downarrow},
\end{equation}
where ``$+$'' refers to $s={\uparrow}$ and ``$-$'' to
$s={\downarrow}$, while
\begin{equation}
\label{spin_j}
\mathcal{H}_{\text{spin},j}^{\pm}
= \pm\gamma_{j}^{z} {\bm \tau}_{z} + \sum_{\nu=x,y,z} 
  B_{j}^{\nu} {\bm \tau}^{j}_{\nu} ,
\end{equation}
determines the state of the $j$th bath spin.  The Hamiltonian
\eqref{spin_j} is readily diagonalized and its eigenenergies
$-\Lambda_{j,\pm}$ and $\Lambda_{j,\pm}$ are determined by
\begin{equation}\label{Lambdaj}
\Lambda_{j\pm}
= \sqrt{(B_{j}^{x})^2 + (B_{j}^{y})^2 + (B_{j}^{z} \pm \gamma_{j}^{z} )^2 }.
\end{equation}
For the evaluation of the Landau-Zener transition probability, we will
not need an explicit expression for the ground states
$|0_{j,\pm}\rangle$ of $\mathcal{H}_{\text{spin},j}^{\pm}$. It
suffices to know that the ground states satisfy the eigenvalue
equation
$\mathcal{H}_{\text{spin},j}^{\pm}|0_{j\pm}\rangle =
-\Lambda_{j\pm}|0_{j\pm}\rangle$.  Consequently,
\begin{subequations}
\label{spin_gs_expectation}
\begin{eqnarray}
\langle 0_{j+}|{\bm\tau}^{j }_{x} |0_{j+}\rangle 
&=& - B_{j}^{x}/\Lambda_{j+},\\
\langle 0_{j+}|{\bm\tau}^{j}_{y} |0_{j+}\rangle
&=& -B_{j}^{y}/\Lambda_{j+},\\
\langle 0_{j+}|{\bm\tau}^{j }_{z} |0_{j+}\rangle
&=& -(B_{j}^{z} + \gamma_{j}^{z})/\Lambda_{j+}.
\end{eqnarray}
\end{subequations}
Since the bath spins do not interact with each other,
the diabatic ground state $|\bm 0_{\pm}\rangle$ is the direct
product of the states $|0_{\pm}\rangle$.

With this ground state defined, we are now in the position to employ
formula~\eqref{PLZopen}.  Inserting relations~\eqref{spin_gs_expectation}
into \eqref{PLZopen_W}, we obtain
\begin{equation}
\label{W2spinbath}
\begin{split}
W^2
={}& \Delta^{2}
   - 4\Delta \sum_j (\gamma_{j}^x)^2\frac{B_{j}^x}{\Lambda_{j+}}
   + 4\sum_j [(\gamma_{j}^x)^2+(\gamma_{j}^y)^2]
\\
& +4\sum_{j\neq j'}\Big(
    \gamma_{j}^x\gamma_{j'}^x\frac{B_{j}^xB_{j'}^x}{\Lambda_{j+}\Lambda_{j'+}}
   +\gamma_{j}^y\gamma_{j'}^y\frac{B_{j}^yB_{j'}^y}{\Lambda_{j+}\Lambda_{j'+}}
   \Big)
\\ &
   + 8\sum_j \gamma_{j}^x\gamma_{j}^y\frac{B_{j}^z}{\Lambda_{j+}}
\,,
\end{split}
\end{equation}
which determines the Landau-Zener transition
probability~\eqref{PLZopen}.
The last term stems from the commutator $[\bm\sigma_x,\bm\sigma_y] =
2i\bm\sigma_z$.
Recall that the general Landau-Zener formula \eqref{PLZopen} was derived
under the assumption that the diabatic qubit-plus-bath ground state at
$t=-\infty$ is non-degenerate. Therefore, our results do not apply to
parameter sets for which any bath spin obeys $B_{j}^{x} =B_{j}^{y}
=(B_{j}^{z}+\gamma_{j}^{z}) =0$ so that $\Lambda_{j+}=0$.

\paragraph{Diagonal basis for spins.}
First, we consider the physically important special case that all
$B_{j}^{x}$ and $B_{j}^{y}$ vanish, so that the eigenvalues become
$\Lambda_{j+} = |B_{j}^{z}+\gamma_{j}^{z}|$.  The ground-state expectation
value $W^2$ which determines the tunnel rate then assumes the more
compact form
\begin{equation}\label{W2sign}
W^{2}
= \Delta^{2} + 4\sum_{j=1}^{N}(\gamma_{j}^{x}+s_{j}
  \gamma_{j}^{y})^{2} ,
\end{equation}
where $s_{j} \equiv \mathop{\text{sign}}(B_{j}^{z}+\gamma_{j}^{z})$.
We can see that in this special case, the transition probability is
independent of the eigenfrequencies $\Lambda_{j}/\hbar$; the
probability essentially depends on the qubit-spin coupling strengths
$\gamma_{j}^{\nu}$.  This frequency independence resembles
the transition probability~\eqref{transverse_result} for the
off-diagonally coupled bosonic oscillator bath, although the qubit-spin
coupling in the present case is not necessarily off-diagonal.

\paragraph{Pure dephasing.}
Interestingly, it follows from Eq.~\eqref{W2sign} that if all
$(\gamma_{j}^{x} + s_{j} \gamma_{j}^{y})$ vanish, that then the
tunneling probability $P_{\uparrow \rightarrow \downarrow}$ equals the
standard LZ probability~\eqref{PLZ_isolated}, despite coupling to the
bath.  This includes the important case of pure diagonal coupling
for which $\gamma_{j}^{x}=\gamma_{j}^{y}=0$.  This bath independence
confirms and generalizes the very recent perturbative calculations by
Wan \textit{et al.}\cite{Wan2007a}  for dissipative LZ transitions in
a spin bath, where it was assumed that $\Delta$ is small.  We find the
same exact bath-independent transition probability in case of a
diagonally coupled spin bath as we found earlier in
Sec.~\ref{sectionbosonicbath} for a diagonally coupled bosonic bath.
This striking result is discussed further in Sec.~\ref{sec:nonlinear}.

\paragraph{Robustness under dephasing.}

Without loss of generality, the  energies $B_{j}^{z}$ of the spins can
be chosen positive. For spin baths away from the very-strong coupling
regime, the coupling constants $\gamma_{j}^{z}$ will be much smaller
than the corresponding energies $B_{j}^{z}$, so that
$\mathop{\mathrm{sign}}(B_{j}^{z}+\gamma_{j}^{z}) = 1$. Thus unless
the qubit-spin coupling is very strong, the LZ transition probability
corresponding to Eq.~\eqref{W2sign} is independent of the
$\gamma_{j}^{z}$. Interestingly, we find that this holds true even 
when $\gamma_{j}^{z}\gg \gamma_{j}^{x,y}$, i.e. if dephasing is much 
stronger than relaxation.  

Moreover, in many models for decoherence by spin baths, the
 coupling of the bath to the ${\bm \sigma}^{y}$ operator simply 
 does not occur, in which case Eq.~\eqref{W2sign} reduces to  
\begin{equation}\label{W2spinxz}
W^{2} = \Delta^{2} + 4\sum_{j=1}^{N}(\gamma_{j}^{x})^{2}.
\end{equation}  
Note that this result is obtained without making  any assumptions on
the dephasing strengths $\gamma_{j}^{z}$.  In other words,  the LZ
transition probability only depends on the integrated spectral density
for relaxation, and is {\em fully independent} of the dephasing
strength of the spin bath. As discussed in Sec.~\ref{sec:experiments},
this is important in experiments. 

\section{Universality of bath-independent nonadiabatic tunneling probability}
\label{sec:nonlinear}

As was found in the previous section, when the qubit interacts with a
bosonic bath or with a spin-bath only via the operator ${\bm
\sigma}_z$, and the total system starts in its ground state, then the
probabilities for dissipative Landau-Zener transitions coincide with
the standard tunneling probability~\eqref{PLZ_isolated}.  Thus, a
natural question arises: Is this a coincidence, a speciality of the
two baths that we studied, or does it hold more generally?
We will now show that this holds in general,
regardless of the specifics of the environment.
Consider the Hamiltonian
\begin{equation}
\mathcal{H}
= \frac{vt}{2} {\bm \sigma}_{z} + \frac{\Delta}{2} {\bm \sigma}_{x}
  + {\bm\sigma}_{z} \mathcal{X}^{z} + \mathcal{H}_{\rm env}.
\end{equation} 
which is the general Hamiltonian~\eqref{LZbath} specified for diagonal
coupling ($\mathcal{X}^{x}=\mathcal{X}^{y}=0$). As before in
Sec.~\ref{Secnonisolated}, the operator $\mathcal{X}^{z}$ is the
environment operator with which the environment is coupled to the qubit,
and $\mathcal{H}_{\rm env}$ is the environment Hamiltonian. Further
specifications of these operators need not be given for our reasoning.
Note that the qubit-bath interaction ${\bm \sigma}_{z}
\mathcal{X}^{z}$ does not commute with the standard LZ Hamiltonian at
any time if $\Delta \ne 0$. Taking this at face value, it is tempting
to assume that the transition probability $P_{\uparrow \rightarrow
\downarrow}$ will be affected by a bath that causes pure dephasing.
Our analysis, however, will reveal that this is not the case.

As before, a complete set of diabatic qubit-plus-bath states
$|{\uparrow} k_{+}\rangle$ and $|{\downarrow} k_{-}\rangle$ can be
found, where the shifted diabatic bath states $|{k_{\pm}}\rangle$ are
eigenstates of $(\mathcal{H}_{\rm env} \pm \mathcal{X}^{z} )$.
We assume that $( \mathcal{H}_{\rm env} + \mathcal{X}^{z})$ has a
nondegenerate ground state  $|{0_{+}}\rangle$. Now since the bath does
not induce bit flips, the bit-flip operator~\eqref{BFinteraction}
simply reduces to the \textit{internal}
bit-flip interaction, i.e.\ $\mathcal{V}=\Delta {\bm \sigma}_{x}/2$.
Hence $W^{2} = \Delta^{2}$.  Thus from
Eq.~\eqref{PLZopen} we find the standard Landau-Zener transition
probability for a qubit diagonally coupled to an arbitrary bath at
zero temperature, 
\begin{equation} \label{PLZuniversal}
P_{\uparrow\to\downarrow} =
1 - \exp 
\left( -\frac{\pi \Delta^2}{2 \hbar v}\right).
\end{equation}
It is truly remarkable that diagonal coupling to the environment does
not affect the final transition probability, whatever the nature of
the environment and however strong the coupling operator
$\mathcal{X}^{z}$ may be. In other words, this zero-temperature
transition probability for  diagonal bath
coupling~\eqref{PLZuniversal} indeed holds universally.  It may have
been simple to derive it from Eq.~\eqref{PLZopen}, but the physical
implications are most important:
Landau-Zener transitions are insensitive to pure dephasing at zero
temperature, irrespective of the nature of the bath or of the 
bath-coupling operator $\mathcal{X}^{z}$.

\begin{figure}[t]
\centerline{\includegraphics{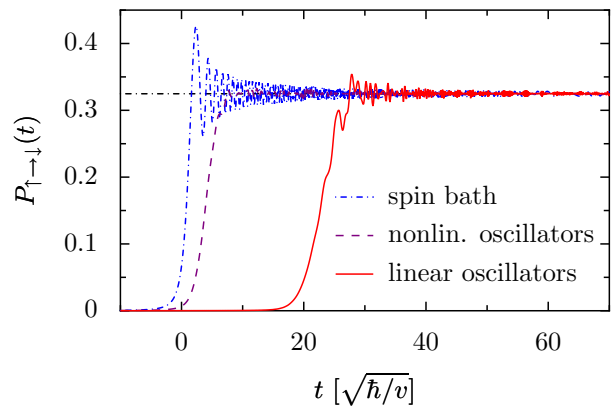}}
\caption{(Color online) Time evolution of spin-flip probability for a qubit with
$\Delta=0.5\sqrt{\hbar v}$ diagonally coupled to two harmonic
oscillators, two nonlinear oscillators, and seven spins, respectively.
The harmonic oscillators are specified by $\Omega_1= 0.1 \sqrt{v
/\hbar}$, $\Omega_2 =0.5\sqrt{v /\hbar}$, $\gamma_1 =2\sqrt{\hbar v}$,
and $\gamma_2=6\sqrt{\hbar v}$, while the nonlinear oscillators, in
addition, have $\beta_1=\beta_2=3\sqrt{\hbar v}$.  The values of the
$B_{j}^{\nu}$ and the $\gamma_{j}^{z}$ are randomly chosen from the
range $[-\sqrt{\hbar v}/10 , \sqrt{\hbar v}/10 ]$.  In all three
cases, the transition probability converges to the universal
value~\eqref{PLZuniversal}.}
\label{fig3}
\end{figure}

To illustrate the universality that we just found, let us now also
consider the coupling to a collection of \textit{nonlinear} oscillators
\begin{equation}
\mathcal{H}_{\rm env}
= \sum_{j=1}^{N} \hbar\Omega_{j} b_{j}^{\dagger}b_{j} + 
  \frac{\beta_{j}}{4 !} ( b_{j}^{\dagger} + b_{j} )^{4}
\end{equation}
that couples to the qubit via the interaction operator
\begin{equation}
\mathcal{X}^{z} = \sum_{j=1}^{N} \frac{\gamma_{j}}{2} (b_{j}^{\dagger}+b_{j}).
\end{equation}
LZ sweeps pas resonances of nonlinear oscillators are of practical interest,
since nonlinear oscillators are currently used for the readout of flux qubits.\cite{Lupascu2006a}
For a numerical test of the predicted final transition probability, 
we take the situation in which
the qubit is diagonally coupled to two of these nonlinear oscillators.
Figure~\ref{fig3} shows the corresponding time-evolution of the
probability $P_{\uparrow\rightarrow\downarrow}(t)$ for the qubit to be
in the ``down'' state.  It also shows the dynamics in case
the qubit couples to two linear oscillators with the same parameters,
except that now $\beta_{1,2}=0$. Furthermore, the effect of a
diagonally coupled spin bath, a special case of Eq.~\eqref{spin_bath},
is also shown. We consider the case in which the spin bath consists of
seven spins.  The internal interaction $\Delta$ has the same value for
all curves in Fig.~\ref{fig3}.  As shown in the figure, at
intermediate times the transition probability shows variations
depending on the specifics of the environments, but the final
probabilities indeed all converge to the universal value~\eqref{PLZuniversal}.

\section{Experimental relevance}
\label{sec:experiments}

Landau-Zener transitions are commonly used in experiments to determine
the interaction $\Delta$ between diabatic energy
levels.\cite{Wernsdorfer1999a} One usually
takes the standard LZ formula~\eqref{PLZ_isolated} as a starting
point. For fixed sweep speed $v$ of the energy levels, the only
unknown in the equation is the internal interaction $\Delta$. By
varying $v$, one can determine the Landau-Zener probability
$P_{{\uparrow}\to{\uparrow}}$ and, hence, the $\Delta$. This in turn
is a method to test the validity of the two-level
result~\eqref{PLZ_isolated}. 

However, if the qubit is coupled to an environment that causes
relaxation, then Eq.~\eqref{transverse_result} shows that one actually
measures $\sqrt{\Delta^{2} +  S}$ for a qubit in an oscillator bath,
where $S$ denotes the integrated spectral density.  Variation of the
sweep speed $v$ does not help in extracting $\Delta$ and $S$
separately from a Landau-Zener experiment. 
If the bath only couples diagonally and thereby causes only pure
dephasing, then at zero temperature, the bath does not influence the
Landau-Zener tunneling probability at all.

In most situations, the environment will cause both relaxation and
pure dephasing. We find that dephasing will hardly change the
relaxation-dependent transition probabilities, see
Eqs.~\eqref{Wthetaphi} and \eqref{W2sign}. In many
experiments pure dephasing is much faster than relaxation, in other
words $T_{2}^{*}$ times are much shorter than $T_{1}$
times. It will therefore often occur that $S
\sin^{2}\vartheta \ll \Delta^{2}$ and, hence, relaxation can be
neglected on the time scale of the Landau-Zener transition. Then our
results imply that with a Landau-Zener experiment, one can accurately
determine the internal interaction $\Delta$ even in the presence of
strong dephasing, such that $S \cos^{2}\vartheta \gg \Delta^{2}$.
Landau-Zener experiments are therefore a surprisingly reliable tool to
determine tunnel splittings. We now 
consider some specific applications.

\subsection{Circuit QED}\label{sec:circuitQED}

Circuit QED \cite{Wallraff2004a, Chiorescu2004a} is a
superconducting-circuit analogue of optical cavity QED:  A charge
qubit\cite{Wallraff2004a} or a flux qubit\cite{Chiorescu2004a} is
coupled so strongly to a a quantized harmonic oscillator in the
circuit that Rabi oscillations can be observed in a solid state
environment.  Because the circuit-QED system is so highly tunable, it
enables the study of quantum dynamics of open quantum systems in new
parameter regimes. 
Recently, we proposed to use LZ transitions in circuit QED to generate
single microwave photons,\cite{Saito2006a} by choosing parameters
such that the qubit-oscillator coupling is off-diagonal and the internal
interaction $\Delta$ vanishes. Here we consider an arbitrary $\Delta$
and discuss the possibility of qubit-oscillator couplings other than
off-diagonal. 

The Hamiltonian describing a charge qubit interacting with the
transmission-line resonator is
\begin{equation}
\mathcal{H}(t) = \mathcal{H}_\text{q}(t) + \mathcal{H}_\text{q-osc}(t)
+ \mathcal{H}_\text{osc},
\end{equation}
where the different terms describe the qubit, the oscillator, and the
qubit-oscillator coupling, respectively.  If the dynamics is
essentially restricted to two states with $N$ and $N+1$ Cooper pairs
in the box, then the Hamiltonian becomes beautifully
simple,\cite{Blais2004a} and in our notation reads
\begin{subequations}\label{cqed_hamiltonian_tunnel}
\begin{eqnarray}
\label{CQED_Hq}
\mathcal{H}_\text{q}
&=& - \frac{E_{\rm J}(t)}{2} {\bm \sigma}_{z}
    + \frac{\Delta}{2} {\bm \sigma}_{x} ,
\\
\mathcal{H}_\text{env}
&=& \hbar \omega_{\rm r} (b^{\dag} b + \frac{1}{2})
   - \frac{\gamma^{2}(1 - 2 N_{\rm g}^{\rm dc})^{2}}{\hbar \omega_{\rm r}},
\\
\label{CQED_Hqo}
\mathcal{H}_\text{q-env}
&=& \gamma (b + b^{\dag}) {\bm \sigma}_{x} ,
\end{eqnarray}
\end{subequations}
with the charging energy $E_\text{C}$, the oscillator frequency
$\omega_\text{r}$, and the coupling strength $\gamma$.
We presented the Hamiltonian in so-called tunneling representation,
which is the basis in which the charge states $|N\rangle$ and
$|N+1\rangle$ are eigenstates of $\bm\sigma_x$.
Since the Josephson link is implemented by a dc SQUID, the Josephson
energy $E_\text{C}(t) = E_{\rm J,max}\cos [ \Phi (t)/\Phi_{0}]$ can be
manipulated upon variation of the flux $\Phi(t)$ that penetrates the
SQUID,\cite{Kleiner2004a} where $\Phi_0$ denotes the flux quantum.
The dimensionless quantity $N_\text{g}^\text{dc}$ is proportional to
an applied gate voltage which can be used to adjust the internal
coupling, which in terms of the control parameters reads $\Delta =
4(1-2N_\text{g}^\text{dc})(\gamma^2/\hbar\omega_\text{r}
-E_\text{C})$.

Since  the internal coupling $\Delta$ is tunable for this setup, one
can use Landau-Zener transitions to determine the integrated spectral
density $S$ for the purely off-diagonal coupling \eqref{CQED_Hqo}.
In the one-oscillator model \eqref{cqed_hamiltonian_tunnel},
$S=\gamma^2$, so that $W^2=\Delta^2+\gamma^2$.  Thus, by Landau-Zener
sweeps past the oscillator resonance for different values of
$\Delta$, Eq.~\eqref{transverse_result} allows one to determine the
qubit-oscillator coupling strength $\gamma^{2}$ in independent ways.

The Hamiltonian \eqref{cqed_hamiltonian_tunnel} features a possibly
nonzero internal interaction $\Delta$ between the qubit levels, which
was assumed zero  in the setting of this superconducting circuit for
which single-photon generation was recently
proposed.\cite{Saito2006a} Besides the transitions
$|{\uparrow}0\rangle \rightarrow |{\downarrow}1\rangle$ that generate
a single photon, for $\Delta\ne 0$ there is now also the process
$|{\uparrow}0\rangle \rightarrow |{\downarrow}0\rangle$ which flips
the qubit without changing the cavity state.  Therefore, a reliable
single-photon generation requires that the transitions induced by
$\Delta$ are not relevant, while the qubit-oscillator coupling has to
be sufficiently large,\cite{Saito2006a} such that $\gamma^{2}\gg\hbar
v$. Fortunately, as long as $\Delta \ll \hbar \omega_\text{r}$,
efficient single-photon generation is very well possible by sweeping
the qubit energy only on a finite frequency interval small compared to
$\omega_\text{r}$ but large compared to the width of the cavity
resonance. 

It would be interesting to test in circuit QED our predictions for
other than off-diagonal qubit-bath couplings. This requires magnetic
fields and gate voltages to be swept simultaneously. The problem with
sweeping gate voltages, however, is that this displaces the ground
state of the oscillator,\cite{Blais2004a} which practically means that
$\Delta$ cannot be used as a time-dependent control parameter.
Consequently, the electric-plus-magnetic sweeping in circuit QED falls
outside the class of models that we considered. It seems that
Landau-Zener transitions with a general qubit-oscillator coupling
angle $\vartheta$ as in Eq.~\eqref{BFosc} cannot be engineered in
state-of-the-art circuit QED.    

We already mentioned the width of the cavity resonance, but actually
in the model~\eqref{cqed_hamiltonian_tunnel}, we assumed that the
quality factor of the resonator was infinite. The spectral density of
the oscillator bath then consists of a single delta peak at the cavity
resonance frequency, so that $S = \gamma^{2}$.  For the non-ideal
resonance, the spectral density becomes
\cite{Tian2002a, vanderWal2003a}
\begin{equation}\label{eqJcircuitQED}
J(\omega) = \frac{\alpha \kappa\omega_{r}^{4}\omega}{(\omega - \omega_{r})^{2} + \kappa^{2}\omega_{r}^{2}\omega^{2}},
\end{equation} 
where the dimensionless parameters $\alpha$ and $\kappa$ measure the
qubit-oscillator strength and the width of the resonance peak,
respectively.
For small frequencies $\omega \ll \omega_{r}$, the spectral density
$J(\omega)\simeq \alpha\kappa \omega$, as for an ohmic bath. Now
suppose that we let
the qubit undergo a LZ sweep from frequency zero across the resonator
frequency until a frequency $\omega_{\rm max}> (1+\kappa)\omega_{r}$.
Now since for frequencies $|\omega-\omega_{r}| > \kappa\omega_{r}$ the
spectral density falls off rapidly, a hypothetical continuation of the
LZ sweep from frequency $\omega_{\rm max}$ to infinity would hardly
change the transition probability. Therefore, by combining
Eqs.~\eqref{Sdef} and \eqref{eqJcircuitQED} we find the integrated
spectral density 
\begin{equation}\label{ScircuitQED}
S
= \frac{\alpha(\hbar\omega_{r})^{2}}{4 \pi }\Biggl\{
  \arctan\left(\frac{2 - \kappa^{2}}{\kappa \sqrt{4 - \kappa^{2}}}\right)
  +\frac{\pi}{2}\Biggl\} \simeq \frac{\alpha(\hbar\omega_{r})^{2}}{4 }.
\end{equation}
Inserting this expression into Eq.~\eqref{transverse_result} gives an
accurate value for the cavity-induced LZ transition probability. The
last identity in \eqref{ScircuitQED} holds in the experimentally realized 
good-cavity limit $\kappa \ll 1$. In this limit one can approximate the spectral
density~\eqref{eqJcircuitQED} by a single oscillator with
qubit-oscillator coupling $\gamma^{2}=\alpha(\hbar\omega_{r})^{2}/4$,
reproducing the model~\eqref{cqed_hamiltonian_tunnel}.

This leads us to the important conclusion that LZ transition
probabilities can be computed exactly for qubits swept past narrow or
broad cavity resonances alike.  The total strength $S$ of the
atom-cavity coupling determines the LZ transition probability, and for
weak dissipation $S$ is independent of the scaled cavity width
$\kappa$.  Dissipative Landau-Zener transitions are also robust in
this respect. 

\subsection{Cavity QED and photonic crystals}\label{sec:cavityPC}
Atoms couple off-diagonally to the electromagnetic field, which for many
practical purposes can be considered as a bath at zero temperature. 
Atomic energies can be swept by applying dc electric and
magnetic fields, which give rise to Stark and Zeeman shifts, respectively. 
Atomic resonances are usually narrow but hard to sweep by large percentages.
Artificial atoms are  more tunable, but their resonances are also broader.
Our results apply to situations where resonances can be swept by much more than their width. 

An atom in free space feels a spectral density that is quadratic in
the frequency,  $J(\omega) \propto (\omega/\omega_{0})^{2}$, with no cutoff
frequencies in the optical regime, so that $S\rightarrow \infty$.
During a Landau-Zener sweep of an atom in free space, the atom will
finally have  decayed spontaneously to its ground state. However, the
spectral density felt by the atom, also known as its local optical
density of states,\cite{Wubs2002a} can be engineered by changing its
dielectric environment. In photonic crystals for example, the
periodicity of the refractive index on the scale of an optical
wavelength $\lambda_{0}=2\pi c/\omega_{0}$ may create a photonic band
gap of width. Within this gap, the spectral density
$J(\omega)$ ideally vanishes and in practice it can be strongly
reduced, whereby spontaneous emission by the atom is strongly
suppressed.\cite{Lodahl2004a}

\begin{figure}
\includegraphics{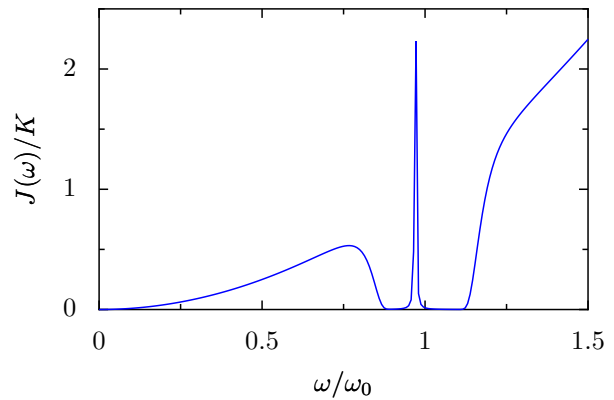}
\caption{(Color online) Sketch of spectral density for an atom near a local defect in
a photonic crystal with a band gap. The quadratic free-space spectral
density is modified by the crystal that creates a spectral gap around
$\omega_{0}$. A narrow defect mode inside a  broader band gap allows a
controlled atom-defect interaction via LZ sweeps of the atomic
transition frequency $\omega_{\rm A}(t)$.  }
\label{fig:Jphotoniccrystal}
\end{figure}

By making a controlled point defect or line defect defect in the
vicinity of the atom that breaks the periodicity of the photonic
crystal, a narrow defect  mode may be created\cite{Ogawa2004a} within
the spectral gap, as sketched in Fig.~\ref{fig:Jphotoniccrystal}. 
Ideally, this would allow cavity QED experiments to be performed within a
photonic crystal, and progress is made in this
direction.\cite{Lodahl2004a,Ogawa2004a} We propose to do LZ sweeps of
the atomic frequency $\omega_{\rm A}(t)$ around the defect frequency
but within the band gap. This will allow the creation of atom-defect
entanglement and of single photons in the defect mode, in quite the
same way as in circuit QED.\cite{Saito2006a}

\subsection{Molecular nanomagnets}\label{sec:nanomagnets}

The energy levels of molecular nanomagnets can be swept by switching
on dc magnetic fields.\cite{Wernsdorfer1999a} Higher-excited states
have higher magnetization and these are excited more when switching
rates are high. These nanomagnets are not intrinsic two-level systems
and indeed many successive LZ transitions are observed upon increasing
the magnetic field.  They can be cooled until tunneling rates become
temperature independent.\cite{Wernsdorfer1999a} In
recent experiments \cite{Wernsdorfer2006a} this happens rather
abruptly at a temperature of $0.6\,{\rm K}$. In this low-temperature
pure quantum tunneling regime, our predictions apply. LZ transitions
are commonly used in experiments to determine level interactions
between energy levels of molecular nanomagnets. 
Energy relaxation due to thermal environment 
usually changes the effective energy gap.\cite{Saito1999a}
But as we showed, LZ transitions for qubits are robust under
dephasing, which on a qualitative level agrees with previous 
lab experience about LZ transitions in nanomagnets.\cite{Wernsdorfer1999a}

Molecular nanomagnets show many avoided crossings as the magnetic
field is varied. These crossings are often well separated, so that one
can ask what type of bath influences the effective-two-qubit dynamics
around a particular level crossing. Neither the bath nor its coupling
to the qubit are known precisely, but at very low temperatures the
main source of decoherence will stem from the coupling to other
electronic and nuclear spins, so that a spin bath seems the
appropriate model.  Our results of Sec.~\ref{sec:spinbath} show in
detail how transition probabilities depend on both the internal
interaction $\Delta$ and the integrated spectral density $S$.
Although strong dephasing is probable in nanomagnets, our
analytical results prove that this hardly affects the LZ transition
probabilities in the experimentally accessible regime where tunneling
rates are temperature independent.\cite{Wernsdorfer2006a} 

A theory for {\em multiple} dissipative LZ transitions in
molecular magnets was developed by Leuenberger and
Loss.\cite{Leuenberger2000a} This theory presumes that in between
successive LZ transitions, all quantum coherences have been washed out
due to dephasing while dephasing {\em during} the individual LZ
transitions does not influence the transition probabilities. This
assumption is rather natural if the LZ tunneling time $\tau_{\rm LZ}$
is much shorter than the pure dephasing time, but obviously requires
proof for slower LZ transitions.  This is an important application of
our results: our derivation of the universal bath independence of
individual LZ transitions under dephasing proves that the theory of
Ref.~\onlinecite{Leuenberger2000a} is  more widely applicable than one
would have guessed previously.   
\subsection{Quantum computer as a spin-bath simulator}\label{sec:simulator}
An ideal quantum computer is a collection of qubits whose energies and
internal interactions are tunable, and whose mutual interactions are
tunable as well. One method for single-qubit manipulation is a
Landau-Zener sweep. This has been realized in recent experiments on
superconducting qubits.\cite{Izmalkov2004a, Oliver2005a,Sillanpaa2006b} 
If the interactions with the other qubits are not
exactly zero, then our spin-bath result~\eqref{W2spinbath} predicts an
effect on the ``single-qubit'' LZ transition probability. This way,
the LZ sweep provides a test of the  settings of the quantum computer
in the operational space of the qubit.

On the other hand, our predictions can be tested very carefully by
controlled variation of qubit-qubit interactions in the quantum
computer.  Indeed, the quantum computer could be seen as a quantum
simulator, in our case of the effects of a spin bath on the LZ transition
probability. A full-fledged quantum simulator does not exist yet, but recent
experiments on a system of four superconducting qubits with tunable
couplings look promising.\cite{Grajcar2006a}   

\subsection{Superconducting qubits and their spin baths}\label{sec:phasequbits}
Josephson phase qubits at very low
temperatures ($ 20\,{\rm mK)}$ exhibit decoherence mainly due to
interactions with two-level microwave resonators.\cite{Simmonds2004a}
Since it is found that qubit losses strongly depend on driving amplitudes, these
resonators cannot be described as a bosonic bath. Rather, they are
nowadays thought to be charge two-level systems.\cite{Martinis2005a}
The existence of these resonators has important implications for the
operation of superconducting qubits. Spectroscopic measurements have
shown that at fixed energy, the qubit often resonantly interacts
with only a single resonator. Moreover, coherent quantum oscillations
between a qubit and such a single resonator were observed.
\cite{Cooper2004a} Remarkably,  decoherence times of these
microscopic two-level systems are larger than that of the qubit. This
in turn has led to the very recent proposal to use instead these
microscopic resonators as qubits for quantum information processing.\cite{Zagoskin2006a}

Our exact result~\eqref{W2spinbath} for the
Landau-Zener transition probability of a qubit in a spin bath in the
low-temperature tunneling regime can be an important tool for
analyzing further the properties of the microscopic resonators and
their couplings to the qubit. Our assumption of a spin-star
configuration, i.e. a bath of mutually non-interacting spins, is
probably correct for the spin bath of the phase qubits.\cite{Zagoskin2006a}  
Monte-Carlo simulations indicate that narrower
qubit-spin resonances are shadowed by larger ones.\cite{Martinis2005a} 
For that reason it is important that
formula~\eqref{W2spinbath} holds generally, whether the qubit
resonantly interacts with one microwave resonator at a time or not. 

During a so-called fast-pulse measurement of the state of the qubit,
the qubit energy moves in and out of resonance with many of these
resonators.\cite{Cooper2004a} As a consequence, the resonators reduce
the fidelity of the measurement. Actually, this effect of the
fluctuators has already been estimated in terms of multiple LZ
transitions in Ref.~\onlinecite{Cooper2004a}. There it was assumed that the
resonators couple off-diagonally via $\sum_{j=1}^{N}\gamma_{j}^{x}{\bm
\sigma}_{x}{\bm \tau}_{x}^{j}$, with coupling strengths
$(\gamma_{j}^{x})^{2}$ given by the size of the splittings as measured
in the qubit spectroscopy. Our results~\eqref{W2spinbath} and
especially~\eqref{W2spinxz} show that the formula used in
Ref.~\onlinecite{Cooper2004a} becomes exact for a zero-temperature
spin bath that couples off-diagonally, even in case of overlapping
spurious resonances.\cite{Martinis2005a} 
It is unfortunate that microscopic resonators reduce the fidelity of
the fast-pulse readout method, but it is good to know how much.  More
good news is that the same fidelity reduction would be obtained even
if there would be additional dephasing by the spins.

Landau-Zener sweeps for phase qubits can be relevant for one more
reason:\cite{Wubs2006a} the precise spectral distribution of the spins
will be sample dependent. Moreover, the distribution for a single
sample varies on a time scale of days.\cite{Zagoskin2006a}
Fortunately, the LZ sweep measures a ``global'' property of the spin
bath, namely its integrated spectral density $S$. It is fair to assume
that $S$ will vary less from sample to sample and from one day to the
next than $J(\omega)$ at a fixed frequency.  We therefore suggest to
use LZ sweeps as a robust way of ``gauging'' and characterizing the
spin-bath environment of superconducting phase qubits. 

\section{Summary and conclusions}
We studied the effect of various zero-temperature environments on the
Landau-Zener transition probability of a qubit.  The main result of
this paper is the corresponding generalization of the bit-flip
transition.  The mathematical form of this result, just like the
standard Landau-Zener formula, is charmingly simple: from the total
Hamiltonian one identifies the bit-flip operator $\mathcal{V}$ of the
qubit and computes the expectation value $W^2=4\langle \mathcal{V}^{2}
\rangle$ for the initial diabatic ground state of the qubit plus
its environment. The transition probability follows formally by
replacing in the standard expression for the two-level LZ problem
the factor $\Delta^2$ by $W^2$.
Following this recipe, we have calculated in Sec.~\ref{sec:bath} the
LZ tunneling probabilities for oscillator baths and for spin baths,
which represent the two most important environments in quantum
dissipation research.  These examples illustrated that dissipative
Landau-Zener transition probabilities in general depend on the type of
environment and on the way the qubit couples to it.

However, when the qubit-bath coupling is of the diagonal type, causing
so-called pure dephasing, then the tunneling probability coincides with the
original LZ probability, regardless of the details of the environment.
We expect that this universal behavior is observable and important for
a wide variety of applications. A bath-independence of another kind
was found for a qubit swept past a broadened (circuit) cavity
resonance: the transition probability turned out to be independent of
the quality factor of the cavity. Since transition probabilities are also
independent of the environment parameters in some phenomenological
models with non-Hermitian dynamics,\cite{Akulin1992a,Schilling2006a}
possible mappings between these and our models deserve future studies.

For the experimentally important hybrid situation in which the bath
causes both relaxation and dephasing, it was found that the influence
of dephasing is negligible, unless the qubit-bath coupling is
exceptionally strong.  This robustness of the LZ transition
probability under dephasing is quite surprising and important in
applications. 
Furthermore, the application of our results to experiments in
superconducting circuits seems very promising, for example for circuit
QED and for the fast-pulse readout of phase qubits.

In the future, it will be interesting to clarify whether a degeneracy
of the ground state modifies the results.  The effects of finite temperatures
on the LZ tunneling would also deserve thorough investigations in the
light of the exact zero-temperature results, with the recent work by
Pokrovsky and Sun\cite{Pokrovsky2007a} as an important first step.
Other experimentally relevant issues are how nonlinear sweeping and
finite sweeping times affect the dissipative transition. 

\section*{Acknowledgements}
MW likes to thank M. H. S. Amin for fruitful discussions.
This work has been supported by DFG through SFB\,484 and SFB\,631, and by a
Grant-in-Aid from the Ministry of Education, Sciences, Sports, Culture
and Technology of Japan (No.\ 18540323).
SK and PH acknowledge support through ``Nanosystems Initiative Munich''.

\appendix
\section{Landau-Zener transition in a multi-level system}
\label{app:BEgen}

\begin{figure}
\includegraphics{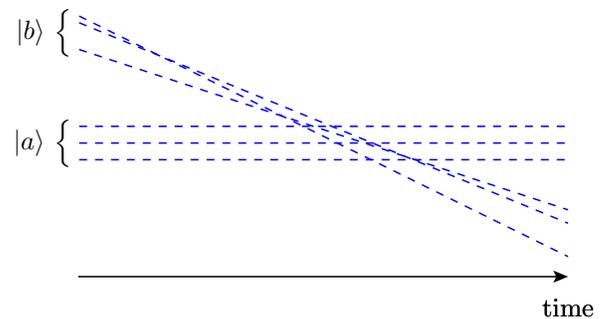}
\caption{(Color online) Crossing of two groups of diabatic states: states $|a\rangle$
whose energy is time-independent and states $|b\rangle$ whose energies
are reduced with constant velocities $v_b$.}
\label{fig:BE}
\end{figure}

We consider the multi-level Hamiltonian
\begin{equation}
\label{app:H}
\begin{split}
H(t)
= & \sum_{a} \varepsilon_a |a\rangle\langle a|
  + \sum_{b} \big(\varepsilon_b-v_bt) |b\rangle\langle b|
\\
  & + \sum_{a,b}(X_{ab}|a\rangle\langle b|+X_{ab}^*|b\rangle\langle a|\big)
\end{split}
\end{equation}
with the time-dependent diabatic energies sketched in
Fig.~\ref{fig:BE}.  We assume that all diabatic states $|a\rangle$,
$|b\rangle$ are mutually orthogonal and that all $v_b>0$.
In the limit $t\to\pm\infty$, the states $|a\rangle$, $|b\rangle$
become eigenstates of the Hamiltonian \eqref{app:H}.  The off-diagonal
part of the Hamiltonian is such that it only couples states of different
groups while states within one group are uncoupled.
Without loss of generality, we assumed in \eqref{app:H} that the
diabatic energies of the $a$-states are time-independent.  If they all
had an identical velocity $v_a$ smaller than all $v_b$,
we could obtain the Hamiltonian \eqref{app:H} by a gauge
transformation with the time-dependent phase factor $\exp(-iv_a
t^2/2\hbar)$.  Then $v_b$ becomes the velocity of the $b$-states with
respect to $v_a$.

We choose as an initial condition that the system starts at $t=-\infty$
in one particular state $|a\rangle$.  The central quantity of interest
is  the probability $P_{a\to a'}$ for a nonadiabatic Landau-Zener
transition to a state $|a'\rangle$ at $t=\infty$.

It is convenient to work in an interaction picture with respect to the
diagonal part of the Hamiltonian \eqref{app:H} and, thus, to apply the
unitary transformation
\begin{equation}
U_0(t) = \sum_a e^{-i\varepsilon_a t/\hbar}|a\rangle\langle a|
+ \sum_b e^{-i\varepsilon_bt/\hbar+iv_bt^2/2\hbar} |b\rangle\langle b| .
\end{equation}
Then we have to deal with the interaction picture Hamiltonian
\begin{equation}
\tilde H(t)
= \sum_{a,b} e^{i(\varepsilon_a-\varepsilon_b)t/\hbar +iv_bt^2/2\hbar}
  X_{ab} |a\rangle\langle b| .
+ \text{H.c.}
\end{equation}
In the interaction representation, the nonadiabatic transition
probability reads
\begin{equation}
\label{app:P}
P_{a\to a'} = |\langle a'|S|a\rangle|^2 ,
\end{equation}
where the $S$-matrix is given by the usual time-ordered exponential
\begin{equation}
\label{app:series}
S = \sum_{k=0}^\infty \left(-\frac{i}{\hbar}\right)^{2k} S_k
\end{equation}
with the $(2k)$th order contribution
\begin{equation}
S_k = \int_{-\infty}^{\infty} dt_1 \int_{t_1}^\infty dt_2 \cdots
\int_{t_{2k-1}}^\infty dt_{2k}\, \tilde H(t_{2k})\ldots \tilde H(t_1) .
\end{equation}
Note that the interaction-picture Hamiltonian $\tilde H(t)$ always
flips between states $|a\rangle$ and $|b\rangle$, so that the
perturbation series for the transition amplitude in \eqref{app:P}
only consists of even powers of the interaction $X_{ab}$.

In order to compute the matrix element $\langle a'|S_k|a\rangle$,
we will generalize and simplify our reasoning of
Refs.~\onlinecite{Saito2006a} and \onlinecite{Wubs2006a} for this more
general model.  To begin with, 
we insert $2k-1$ times the unit operator $\sum_a|a\rangle\langle a| +
\sum_b|b\rangle\langle b|$ and obtain
\begin{equation}
\label{app:series1}
\begin{split}
\sum_{a_1\ldots a_{k-1}} & \sum_{b_1\ldots b_k}
X_{ab_k}X_{a_{k-1}b_k}^* X_{a_{k-1}b_{k-1}}\ldots X_{a_1b_1}X_{ab_1}^*
\\
\times
\int_{-\infty}^\infty & dt_1\ldots \int_{t_{2k-1}}^\infty dt_{2k}
\exp\Big\{
  \frac{i}{\hbar}(\varepsilon_{b_1}-\varepsilon_a)t_1
\\
 +\frac{i}{\hbar}& \sum_{\ell=2}^k (\varepsilon_{b_\ell}-\varepsilon_{a_{\ell-1}})t_{2\ell-1}
 +\frac{i}{\hbar}\sum_{\ell=1}^{k-1} (\varepsilon_{a_\ell}-\varepsilon_{b_\ell})t_{2\ell}
\\
 +\frac{i}{\hbar}& (\varepsilon_{a'}-\varepsilon_{b_k})t_{2k}
 + \frac{i}{2\hbar}\sum_{\ell=1}^k v_{b_\ell}(t_{2\ell}^2-t_{2\ell-1}^2)
\Big\} .
\end{split}\end{equation}

\begin{figure}
\includegraphics{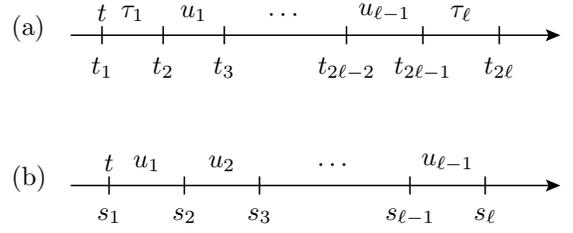}
\caption{Relation between the various time variables.}
\label{fig:times}
\end{figure}

In order to cope with the time-ordering, we substitute the time
variables $t_1,\ldots,t_{2k}$ by the time differences $\tau_\ell =
t_{2\ell}-t_{2\ell-1}$ and $u_\ell=t_{2\ell+1}-t_{2\ell}$ as sketched
in Fig.~\ref{fig:times}a, i.e.\ we set
\begin{subequations}
\begin{align}
t_1 ={}& t
\\
t_{2\ell} ={}& t+\sum_{\ell'=1}^\ell\tau_{\ell'}+\sum_{\ell'=1}^{\ell-1} u_{\ell'}
\\
t_{2\ell+1} ={}& t+\sum_{\ell'=1}^\ell (\tau_{\ell'}+u_{\ell'})
\end{align}
\end{subequations}
which is equivalent to $\tau_\ell = t_{2\ell}-t_{2\ell-1}$ and
$u_\ell=t_{2\ell+1}-t_{2\ell}$.
Note that the Jacobian of this substitution is 1.
Then the multiple time integral in expression~\eqref{app:series1} becomes
\begin{equation}
\label{app:time.int1}
\begin{split}
\int_{-\infty}^\infty dt &
\int_0^\infty du_1\ldots du_{k-1}
\int_0^\infty d\tau_1\ldots d\tau_k
\\
\exp\Big\{ &
  \frac{i}{\hbar}\Big(\varepsilon_{a'}- \varepsilon_a
          +\sum_{\ell=1}^k v_{b_\ell}\tau_\ell \Big)t
\\
  +&\frac{i}{\hbar}\sum_{\ell=1}^k(\varepsilon_{a'}-\varepsilon_{b_\ell})\tau_\ell
  +\frac{i}{\hbar}\sum_{\ell=1}^{k-1}(\varepsilon_{a'}-\varepsilon_{a_\ell})u_\ell
\\
 +&\frac{i}{\hbar}\sum_{\ell=1}^k 
\Big(
      \frac{1}{2} v_{b_\ell}\tau_\ell^2
      + v_{b_\ell}\tau_\ell \sum_{\ell'=1}^{\ell-1}(\tau_{\ell'}+u_{\ell'})
   \Big)
\Big\} .
\end{split}\end{equation}
%
Performing the $t$-integration, we obtain the Dirac delta
\begin{equation}
\label{app:delta}
2\pi\hbar\delta\big(\varepsilon_{a'}-\varepsilon_a+\sum_{\ell=1}^k
v_{\beta_\ell}\tau_\ell\big) .
\end{equation}
Because all $v_b>0$ and the integration interval for each $\tau_\ell$
is $[0,\infty)$, the sum in the argument of the delta function is
non-negative. This has two important implications:

First, the energies of the final and the initial state must fulfill
the relation $\varepsilon_{a'}\leq\varepsilon_a$, which means that all
states $|a'\rangle$ with an adiabatic energy higher than that of the
initial state $|a\rangle$ are finally unoccupied, or
\begin{equation}
\label{app:nogoup}
P_{a\to a'} = 0 \quad\text{for $\varepsilon_{a'}>\varepsilon_a$}.
\end{equation}
We call this no-go theorem~\cite{Sinitsyn2004a, Volkov2005a} the ``no-go-up'' theorem. 
As a corollary, we find that if the system starts at $t=-\infty$ in
the adiabatic ground state, the final state will be either the initial
state or one of the states $|b\rangle$.

If all avoided crossings of the adiabatic energies are sufficiently
narrow, the no-go-up theorem can be understood by semiclassical
considerations: At each avoided crossing, the population splits up
into a coherent superposition of two branches.  If the system starts out in
an $a$-state which has velocity $v_a=0$, it can only choose between
staying in an $a$-state with constant energy or an $b$-state with
decreasing energy.  Thus transitions to diabatic states with higher
energies are impossible.  Note that the validity of these semiclassical 
arguments is limited while the no-go-up theorem~\eqref{app:nogoup} is an 
exact statement that holds for any width of the avoided crossings.

The second implication of Eq.~\eqref{app:delta} concerns the special
transitions $|a\rangle\to|a\rangle$, to which the remainder of this
Appendix is devoted.  Since in this case, the sum in the delta
function must vanish, any contribution to the corresponding transition
probability must come from the subspace
$\tau_1=\tau_2=\ldots=\tau_k=0$.  This means that the $\tau$-integrals
in expression~\eqref{app:time.int1} vanish unless the $u$-integrals
are singular for $\tau_\ell=0$.
This finding can be exploited for the simplification of the
$u$-integrals.  In the present form, however, the twofold summation in
the last term of the exponent in expression~\eqref{app:time.int1} will
complicate this task. Therefore it is convenient to substitute the
integration variables $u_1,\ldots,u_{k-1}$ by the times
\begin{equation}
s_\ell = t+\sum_{\ell'=1}^{\ell-1} u_{\ell'} ,
\end{equation}
which means $u_\ell = s_{\ell+1}-s_\ell$ as sketched in
Fig.~\ref{fig:times}b.
Then we obtain the partially time-ordered integral
\begin{equation}
\label{app:time.int2}
\begin{split}
\int_{-\infty}^\infty ds_1 &
\int_{s_1}^\infty ds_2\ldots \int_{s_{k-1}}^\infty ds_{k}
\int_0^\infty d\tau_1\ldots d\tau_k
\\
\exp\Big\{ &
  \frac{i}{\hbar}(\varepsilon_{a'}- \varepsilon_{a_{k-1}})s_k
  +\frac{i}{\hbar}\sum_{\ell=2}^{k-1}(\varepsilon_{a_\ell}-\varepsilon_{a_{\ell-1}})s_\ell
\\
  +&\frac{i}{\hbar}(\varepsilon_{a_1}- \varepsilon_{a})s_1
  +\frac{i}{\hbar}\sum_{\ell=1}^k v_{b_\ell}\tau_\ell s_\ell
\\
  -&\frac{i}{\hbar} \sum_{\ell=1}^k \Big(
     \varepsilon_{b_\ell}-\varepsilon_{a'} - \frac{\tau_\ell}{2}
      -\sum_{\ell'=1}^{\ell-1}\tau_{\ell'}
   \Big)\tau_\ell
\Big\} .
\end{split}\end{equation}
Setting $a_k=a'$ and $a_0=a$, one sees that all $s$-integrals are of
the form
\begin{equation}
\label{app:int.s}
\int^\infty ds_\ell\, \exp\{ i(\varepsilon_{a_\ell}-\varepsilon_{a_{\ell-1}}
+ v_{b_\ell}\tau_\ell)s_\ell\} ,
\end{equation}
where the lower integration limit can be finite or $-\infty$.
Evaluating this integral, one finds two types of terms:
The first one is a principal value which is always regular and, thus,
will not contribute to $S_k$.
A second term is proportional to the delta function
$\delta(\varepsilon_{a_\ell} -\varepsilon_{a_{\ell-1}} +v_{b_\ell}\tau_\ell)$.
This will contribute if and only if its
singularity is located at $\tau_\ell=0$, as discussed above.
Therefore, we find
that energies of all participating $a$-states must be identical,
\begin{equation}
\label{app:never.go}
\varepsilon_{a'} = \varepsilon_{a_1} = \ldots = \varepsilon_{a_{k}} =\varepsilon_a .
\end{equation}
In the absence of degeneracies in the spectrum of the $a$-states
finally, the important condition follows that all non-vanishing
contributions to the perturbation series must fulfill the relation
\begin{equation}
\label{app:selectionrule}
a' = a_1 = \ldots = a_k = a \,.
\end{equation}
This selection rule states that the only allowed processes are those
in which the system jumps repeatedly from the initial state to one of
the $b$-states and back.  
%
Note that this selection rule holds only for the contributions to the
{\em final} transition probability at time $t=\infty$.  At intermediate
times, other $a$ states can be populated as well, as has been exemplified
in a numerical study of Landau-Zener transitions of a qubit coupled to
a single harmonic oscillator.\cite{Saito2006a}

By use of the selection rule \eqref{app:selectionrule} and the form
\eqref{app:time.int2} of the multiple integral, we obtain for the
$2k$th order term $\langle a|S_k  |a\rangle$ the expression
\begin{equation}
\label{app:series2}
\begin{split}
&
\sum_{a_1\ldots a_{k-1}} \sum_{b_1\ldots b_k}
|X_{ab_k}|^2\, |X_{ab_{k-1}}|^2 \ldots |X_{ab_1}|^2
\\
& \times \int_{-\infty}^\infty ds_1
  \int_{s_1}^\infty ds_2\ldots \int_{s_{k-1}}^\infty ds_{k} \,
  \int_0^\infty d\tau_1\ldots d\tau_k
 \\ & \times  e^{i\sum_{\ell=1}^{k}v_{b_\ell}\tau_\ell s_\ell/\hbar}
\\ &\times
\exp\Big\{
  -\frac{i}{\hbar} \sum_{\ell=1}^k \Big(
     \varepsilon_{b_\ell}-\varepsilon_a - \frac{\tau_\ell}{2}
      -\sum_{\ell'=1}^{\ell-1}\tau_{\ell'}
   \Big)\tau_\ell
\Big\} .
\end{split}\end{equation}
A most important observation is now that the matrix elements
$X_{ab_\ell}$ no longer depend on the index $a_\ell$. Any permutation
of the integration variables $s_\ell$ can therefore be undone by proper
relabelling.  Thus, we can replace the $s$-integrals in the second
line of Eq.~\eqref{app:series2} by the symmetrized version
\begin{equation}
\frac{1}{k!}\int_{-\infty}^\infty \!\! ds_1\ldots ds_k\,
e^{\frac{i}{\hbar}\sum_{\ell=1}^{k}v_{b_\ell}\tau_\ell s_\ell}
= \frac{(2\pi\hbar)^k}{k!} \frac{\delta(\tau_1)\ldots\delta(\tau_k)}{
  v_{b_1}\ldots v_{b_k}} .
\end{equation}
The remaining $\tau$-integrations can be evaluated as well, each
of which yielding $1/2$, so that finally
\begin{equation}
\langle a|S_k|a\rangle
= \frac{1}{k!}\Big( \pi\hbar \sum_b \frac{|X_{ab}|^2}{v_b}\Big)^k .
\end{equation}
Inserting this into the series~\eqref{app:series}, we obtain the
exact nonadiabatic Landau-Zener transition probability
\begin{equation}
\label{app:Paa}
P_{a\to a}
= \exp\Big( -\frac{2\pi}{\hbar}
         \sum_b \frac{|\langle a|X|b\rangle|^2}{v_b}
      \Big) ,
\end{equation}
where $X=\sum_{a,b}|a\rangle X_{ab}\langle b|+\text{h.c.}$ denotes the
off-diagonal part of the Hamiltonian \eqref{app:H}.  This general
formula generalizes our previous results.\cite{Saito2006a,Wubs2006a} 
Similar formal results, not applied to quantum dissipation, have been 
presented in the very recent work by Volkov and 
Ostrovsky.\cite{Volkov2007a}

Of much practical importance is the case in which all $b$-states have
the same velocity, $v_b=v$, so that we face a situation of two
crossing energy bands.  Owing to $\langle a|X|a'\rangle=0$ for all
$a,a'$, one finds $\sum_b\langle a|X|b\rangle\langle b|X|a\rangle = \langle
a|X^2|a\rangle$. For a nondegenerate initial state $|a\rangle$ we end up with 
\cite{Saito2006a,Wubs2006a} 
\begin{equation}
\label{app:Paa.v}
P_{a\to a}
= \exp\Big(-\frac{2\pi \langle a|X^2|a\rangle}{\hbar v}\Big) .
\end{equation}


\end{document}